# Interface-tuning of ferroelectricity and quadruple-well state in $CuInP_2S_6$ via ferroelectric oxide

Kun Wang[1], Du Li[2], Jia Wang[1], Yifei Hao[1], Hailey Anderson[1], Li Yang[2], Xia Hong[1*]

[1] *Department of Physics and Astronomy & Nebraska Center for Materials and Nanoscience, University of Nebraska-Lincoln, Lincoln, Nebraska 68588-0299, USA.*

[2] *Department of Physics, Washington University in St. Louis, St. Louis, Missouri 63130-4899, USA.*

[*]Email: xia.hong@unl.edu

## ABSTRACT

Ferroelectric van der Waals $CuInP_2S_6$ possesses intriguing quadruple-well states and negative piezoelectricity. Its technological implementation has been impeded by the relatively low Curie temperature (bulk $T_C$ ~42 °C) and the lack of precise domain control. Here we show that $CuInP_2S_6$ can be immune to the finite size effect and exhibits enhanced ferroelectricity, piezoelectricity, and polar alignment in the ultrathin limit when interfaced with ferroelectric oxide $PbZr_{0.2}Ti_{0.8}O_3$ films. Piezoresponse force microscopy studies reveal that the polar domains in thin $CuInP_2S_6$ fully conform to those of underlying $PbZr_{0.2}Ti_{0.8}O_3$, where the piezoelectric coefficient changes sign and increases sharply with reducing thickness. High temperature *in situ* domain imaging points to a significantly enhanced $T_C$ exceeding 200 ºC for 13 nm $CuInP_2S_6$ on $PbZr_{0.2}Ti_{0.8}O_3$. Density functional theory modeling and Monte Carlo simulations show that the enhanced polar alignment and $T_C$ can be attributed to interface-mediated structure distortion in $CuInP_2S_6$. Our study provides an effective material strategy to engineer the polar properties of $CuInP_2S_6$ for flexible nanoelectronic, optoelectronic, and mechanical applications.

Layered van der Waals (vdW) ferroelectric $CuInP_2S_6$ (CIPS) hosts unconventional quadruple-well energy[1] and high ionic conductivity,[2, 3] which lead to various exotic phenomena, including negative piezoelectric coefficient,[4] thickness-dependent in-plane polarization,[5] giant electrostriction enabled strain tunability,[6] strong coupling of ferroelectric polarization with ionic migration and topography variation,[7-9] and the self-rectifying memristor effect.[10] Compared with other vdW ferroelectric semiconductors, the above room temperature Curie temperature (bulk $T_C$ ~ 42 °C),[11] large band gap ($E_g$ ~ 2.9 eV),[12] and large out-of-plane polarization ($P$ ~ 4 $\mu C/cm^2$)[4] make CIPS the most promising candidate for developing flexible and transparent nanoelectronic and optoelectronic devices, such as nonvolatile memory,[11, 13, 14] negative capacitance transistors,[15, 16] memristors,[10, 17] rectifiers,[10, 18] photocatalysis,[19] and photovoltaics.[20]

In CIPS, the ferroelectric polarization mainly originates from the off-center displacement of Cu in the sulfur frames. Due to the mobile Cu ions, the ferroelectric order parameter is highly susceptible to thermal,[7, 21] electrical,[2] and mechanical stimulations.[9, 22] It has been shown that domain writing in CIPS always yields diffusive, rough domain walls (DWs), precluding precise polarization control at the nanoscale.[5, 13] The relatively low $T_C$ and lack of domain control in CIPS can compromise the thermal stability and density of polarization-enabled device applications. To transcend these material limitations, it calls for an effective strategy to promote the polar alignment in CIPS. A promising route is to interface CIPS with a ferroelectric with well-controlled polar states. For example, enhanced piezoelectric response has previously been observed in organolead trihalide thin films prepared on ferroelectric oxide films, where the interfacial coupling aligns the polar liquid state in the hybrid perovskite.[23]

In this study, we report controlled domain formation and enhanced piezoelectricity and ferroelectricity in thin CIPS flakes interfaced with epitaxial ferroelectric $PbZr_{0.2}Ti_{0.8}O_3$ (PZT) films. Piezoresponse force microscopy (PFM) studies show that thin CIPS flakes fully mirror the domain structures of underlying PZT, in contrast to the spontaneously formed mosaic domains observed in flakes prepared on doped Si and Au base layers. The enhanced polar alignment for thin CIPS on PZT is accompanied by a sign change in piezoelectric coefficient $d_{33}$, which increases sharply from 5.2 to 14.8 pm/V as the flake thickness $t_{CIPS}$ is reduced from 21 to 6 nm. Density functional theory (DFT) modeling of $CIPS/PbTiO_3$ (PTO) reveals interface-mediated structure distortion in CIPS, which favors the polarization of CIPS to be antialigned with that of PZT. *In situ* domain imaging via high temperature PFM further shows that $T_C$ in thin CIPS is enhanced to above 200 ºC, which becomes comparable to that of technologically important ferroelectric polymers[24] and lead-free oxide ceramics.[25, 26] The interface-lattice-coupling-enabled $T_C$ enhancement is corroborated by Monte Carlo simulations. Details on the sample preparation, PFM characterizations, and theoretical modeling are provided in Methods. All experiments are conducted at room temperature unless otherwise specified.

**RESULTS AND DISCUSSION**

**Domain formation in CIPS**

We transfer mechanically exfoliated 8-300 nm CIPS flakes on 50 nm epitaxial (001) PZT films (Figure 1a) with pre-patterned domain structures and compare their PFM response with those prepared on doped Si and Au base layers (See Methods). The PZT films are deposited on 10 nm $La_{0.67}Sr_{0.33}MnO_3$ (LSMO) buffered $SrTiO_3$ (STO) substrates and possess out-of-plane polarization (Figure S1a and S2). Figure 1b shows the PFM amplitude and phase response of a bare PZT film versus the bias voltage ($V_{bias}$) with respect to the bottom electrode. The coercive voltages for the polarization up ($P_{up}$) and down ($P_{down}$) states are -1.15 V and 1.25 V, respectively. Atomic force microscopy (AFM) studies show that all three types of base layers possess smooth surfaces with typical root mean square roughness of ≤ 5 Å (Figure S1b-d).

Figure 1c-d shows the PFM phase and amplitude images, respectively, taken on a 13-14 nm CIPS flake (Figure S3a) transferred on a PZT film pre-patterned into the $P_{up}$ state. The CIPS flake exhibits a uniform domain structure, with the phase signal comparable with that of PZT. This is in sharp contrast to the samples prepared on doped Si (Figure 1e-f) and Au (Figure 1g-h) base layers, where CIPS flakes of the same thickness (14 nm, Figure S3b-c) show spontaneously formed long-stripe domains with characteristic widths on the order of 100 nm. The distinct domain structures suggest that the PZT base layer promotes the alignment of polarization in CIPS.

To assess the length scale of the interaction between CIPS and PZT, we pre-pattern on three PZT films a series of rectangular $P_{down}$ domains in a uniformly polarized $P_{up}$ background (Figure 2a) and transfer CIPS flakes of various thicknesses on top (Figure 2b-d, and Figure S4-6), denoted as samples S1-S3. The 13 nm (Figure 2c) and 15 nm CIPS flakes (Figure 2d) form the same domain pattern as that of underlying PZT. For the 18 nm CIPS (Figure 2d), the PFM response on top of the $P_{up}$ domains of PZT is uniform and in phase with that of PZT, while local phase variations start to emerge in the region on top of the $P_{down}$ domains of PZT. The polarization asymmetry is clearly manifested in the 26 nm CIPS, where spontaneous stripe domains similar to those observed in Figure 1e-f only show up in the region on top of the $P_{down}$ domains of PZT. For the 28 nm flake, in contrast, the stripe domain structures form on top of both the $P_{down}$ and $P_{up}$ domains of PZT. For the 55 nm and thicker flakes (Figure 2c), the domain distribution is no longer correlated with the underlying PZT polarization. The evolution of domain structure with $t_{CIPS}$ suggests that the mechanism aligning the polarization between CIPS and PZT occurs at the interface and decays with increasing $t_{CIPS}$. The different domain patterns in the intermediate thickness range (18 nm and 26 nm) also show that the $P_{up}$ state has a slightly larger critical thickness for the transition to bulk-like behavior, inferring that the CIPS polarization can be more effectively aligned by the $P_{up}$ state of PZT than the $P_{down}$ state.

**Enhanced $d_{33}$ in thin CIPS on PZT**

For the 14 nm CIPS/PZT stack, the overall piezoresponse is homogeneous and in phase with that of bare PZT (Figure 1c), while its amplitude is weaker (Figure 1d). We carry out off-resonance PFM ramping measurements to quantify the piezoelectric coefficient $d_{33}$ (see Methods). Figure 3a shows the PFM ramping curves taken on the three CIPS samples shown in Figure 1c-h, where the amplitude signal of all samples exhibits a linear $V_{bias}$-dependence above the instrument signal floor (~0.5 mV). From the slope, we obtain an effective piezoelectric coefficient $d_{33}^{tot}$ of 3.9 pm/V for the CIPS/PZT stack (see Methods). This is much smaller than that of PZT $d_{33}^{PZT} = 39 \pm 2$ pm/V, consistent with the PFM amplitude image (Figure 1d). Since both CIPS and PZT layers contribute to the piezoelectric response, the net piezoresponse is given by:

$$d_{33}^{tot} = \nu_{PZT} \cdot d_{33}^{PZT} \pm \nu_{CIPS} \cdot d_{33}^{CIPS}, \tag{1}$$

where $\nu_{PZT}$ and $\nu_{CIPS}$ are the fractional bias voltage dropped across the PZT and CIPS layers, respectively. The positive (negative) sign applies to the condition that the polarization directions of CIPS and PZT are aligned (antialigned). We then perform finite element analysis to model the voltage distribution (see Methods). Figure 3b-c shows the simulated result for a 14 nm CIPS/50 nm PZT stack assuming dielectric constants of 52 for CIPS[13] and 100 for PZT,[27] which yields $\nu_{PZT}$ = 0.26 and $\nu_{PZT}$ = 0.74. The corresponding piezoelectric response from the PZT layer is $\nu_{PZT} \cdot d_{33}^{PZT}$ = 10.1 pm/V. Using Eq. (1), we deduce the piezoelectric response from CIPS to be -6.2 pm/V. The overall phase of the piezoelectric response for the stack is thus determined by the PZT layer.

Since CIPS has the unconventional quadruple-well ferroelectric states,[1] there are two possible configurations for the polarization of CIPS that lead to the negative piezoelectric response: 1) the polarization is aligned with that of PZT, while $d_{33}^{CIPS}$ is negative; 2) the polarization is antialigned with that of PZT, while $d_{33}^{CIPS}$ is positive. To determine which scenario applies to our system, we perform DFT calculations for a CIPS/PTO heterostructure as a model system (see Methods). Figure 3e-f shows the calculated crystal structures of fully relaxed CIPS on a frozen PTO substrate, with the polarization of PTO fixed in the $P_{down}$ state. We calculate the total energy of the heterostructure with the polarization of CIPS in either the $P_{up}$ (Figure 3e) or the $P_{down}$ state (Figure 3f). The energy of the antialigned configuration is 25 meV lower than that of the aligned configuration (Figure S17). This energy difference can be associated with the lattice distortion of the interfacial CIPS due to the strain imposed by PTO, as revealed by the DFT calculations. For the $P_{up}$ state of CIPS (polarizations antialigned with PZT), all copper ions in the interfacial layer of CIPS lie in the same horizontal plane. In contrast, for the $P_{down}$ state (polarization aligned with PZT), there is a larger variation in Cu ion position, causing higher lattice distortion (Figure S18). The standard deviations of the z-direction Cu displacement are 0.01 Å and 0.15 Å for the $P_{up}$ and $P_{down}$ states of CIPS, respectively. This strain-mediated lattice distortion results in a higher elastic energy cost for the $P_{down}$ state of CIPS, which can account for the preference for antialigned polarization between CIPS and PTO.

Based on this result, we adopt the negative sign in Eq. (1) and obtain a positive $d_{33}^{\text{CIPS}}$ of 8.5 pm/V for the 14 nm CIPS on PZT. As a control study, we also quantify $d_{33}^{\text{CIPS}}$ for CIPS on Si and Au base layers. To determine the sign of $d_{33}^{\text{CIPS}}$, we conduct PFM switching measurements and compare the hysteresis loops with that obtained on bare PZT (Figure S9). The phase switching hysteresis of CIPS on Si and Au is in counterclockwise direction, opposite to that obtained on PZT. We thus conclude $d_{33}^{\text{CIPS}}$ for CIPS on Si and Au is negative, which is consistent with previous reports.[1, 4] Figure 3d compares the $d_{33}^{\text{CIPS}}$ values averaged over different spots for the 14 nm CIPS flakes on PZT, Si, and Au base layers. Compared with the results obtained on CIPS flakes on Si (-2.2±0.2 pm/V) and Au (-1.8±0.2 pm/V), $d_{33}$ for CIPS on PZT (8.0±0.7 pm/V) is not only significantly higher but also changes sign.

The positive $d_{33}^{\text{CIPS}}$ for CIPS on PZT reflects a change in the free energy profile, which can be attributed to the interfacial strain imposed by PZT. For standalone CIPS (Figure 3g), the displacements of Cu for the $P_{\text{up}}$ (+) and $P_{\text{down}}$ (-) states are symmetric and within the vdW layer. When the flake is subjected to an external electric field, the change of the interlayer vdW gap is larger than that of the intralayer separation while in the opposite direction, leading to a negative $d_{33}^{\text{CIPS}}$ in the ground state (GS).[4] In the metastable state (MS), the Cu ion is further displaced and enters the gap between the neighboring vdW layers, leading to a positive $d_{33}^{\text{CIPS}}$. Even though the metastable state energy is lower relative to the ground state, it corresponds to a larger Cu displacement that requires overcoming a slight energy barrier. As a result, a negative $d_{33}^{\text{CIPS}}$ is commonly observed in standalone CIPS. On the other hand, for CIPS on PTO in the $P_{\text{down}}$ state, the strain induced by PTO shifts the lower sulfur ion position towards the Cu ion (Figure 3h). Such distortion leads to a higher elastic energy for the $P_{\text{down}}$ state of CIPS, effectively tilting the quadruple well energy profile. This change not only favors the $P_{\text{up}}$ state, or antialigned polarization between CIPS and PTO, but also suppresses the energy barrier between the +GS and +MS state, making the metastable state accessible. This scenario naturally explains the positive $d_{33}^{CIPS}$ observed in thin CIPS on PZT. Such strain modulation is consistent with previous reports of CIPS.[1, 6, 22] Similar interfacial strain-mediated polarization asymmetry has been widely observed in epitaxial ferroelectric oxide heterostructures.[28]

## Thickness dependence of $d_{33}^{CIPS}$

We systematically examine the piezoelectric response of CIPS flakes in a wide thickness range (8-300 nm) on the three types of base layers (Figure 4). For control samples on Au and Si base layers, $d_{33}^{CIPS}$ remains negative for the entire thickness range. The value is about -10 pm/V for thick flakes (> 100 nm) and gradually decreases with reducing $t_{CIPS}$, reaching about -1 pm/V for 6 nm CIPS on Si and -1.6 pm/V for 13 nm CIPS on Au. The suppression of $d_{33}^{CIPS}$ in thin CIPS has been reported previously[29, 30] and can be attributed to the finite size effect due to the enhanced depolarization in thin ferroelectrics.[31]

In contrast, CIPS on PZT shows an unconventional $t_{CIPS}$-dependence of $d_{33}^{CIPS}$, which can be divided into three distinct regions. For thick flakes (Region I: $t_{CIPS}$ > 50 nm), $d_{33}^{CIPS}$ is about -10 pm/V and does not show apparent thickness dependence, similar to the behavior of thick flakes on Au and Si. In the intermediate thickness region (Region II: 25 nm < $t_{CIPS}$ < 50 nm), there is a significant variation in the $d_{33}^{CIPS}$ values, with both positive and negative $d_{33}^{CIPS}$ observed. For example, $d_{33}^{CIPS}$ extracted on 26 nm CIPS flakes varies from 4.9 pm/V to -11.9 pm/V. For thin flakes (Region III: $t_{CIPS}$ < 25 nm), $d_{33}^{CIPS}$ becomes positive and exhibits a monotonical increase with decreasing $t_{CIPS}$, changing from 5.2 pm/V for 21 nm flake to 14.8 pm/V for 6 nm flake. This result is opposite to the thickness dependence observed in the CIPS samples on Au and Si.

Our result shows that CIPS on PZT is not subjected to the finite size effect. This can also be understood within the scenario of the tilted energy quadruple-well of the interfacial CIPS due to the structure distortion. When CIPS is settled in the metastable state, it corresponds to a larger displacement of the Cu ions, which produces an enhanced, positive $d_{33}^{\text{CIPS}}$. Such structural distortion is mediated by the interface lattice coupling with PZT and thus short-ranged, settling the interfacial layers in the metastable state. For the CIPS layers away from the interface, the strain is relaxed and $d_{33}^{\text{CIPS}}$ gradually recovers the bulk behavior. The overall piezoresponse of a CIPS sample is determined by the net effect of the strained interfacial layers in the metastable state ($d_{33}^{\text{MS}} > 0$) and the relaxed layers in the ground state ($d_{33}^{\text{GS}} < 0$), which can be described by $d_{33}^{\text{CIPS}} = \nu_{\text{MS}} \cdot d_{33}^{\text{MS}} + \nu_{\text{GS}} \cdot d_{33}^{\text{GS}}$, with $\nu_{\text{MS}}$ and $\nu_{\text{GS}}$ the fractional bias voltage across the layers in the metastable and ground states, respectively. With increasing $t_{\text{CIPS}}$, the relaxed layers account for a larger fraction across the sample, leading to a larger $\nu_{\text{GS}}$. As a result, the measured $d_{33}^{\text{CIPS}}$ shows a continuous, monotonic decrease with increasing $t_{\text{CIPS}}$, as observed in Region III. Since strain relaxation can be affected by local boundary conditions and defect states, the critical layer number for the transition from the metastable state to the ground state can vary from sample to sample, which accounts for the large variation in the magnitude and even the sign of $d_{33}^{\text{CIPS}}$ for CIPS on PZT in Region II. When $t_{\text{CIPS}}$ is further increased to Region I, CIPS layers in the ground state dominate the piezoelectric response, and the net $d_{33}$ becomes negative (Region I). We emphasize that consistent thickness-dependences for the evolution of domain structures (Figure 2 and Figure S5) and $d_{33}^{\text{CIPS}}$ (Figure 4) have been observed in multiple samples with various CIPS sizes and thicknesses, including isolated flakes with uniform thickness and large flakes with a thickness distribution. It is conceivable that the flakes with different sizes and thickness distributions may be subjected to slight variation in mechanical conditions during the exfoliation and transfer procedure. The fact that we observe highly reproducible PFM results clearly demonstrates that the interface-mediated effect is robust and universal.

**Enhanced Curie Temperature for CIPS**

As the structure distortion in CIPS modifies the ferroelectric free energy profile, it also affects $T_C$. To quantify this effect, we perform *in situ* PFM imaging on sample S1 (Figure 2c) at elevated temperatures (up to 225 °C, see Methods), which are well below the $T_C$ of epitaxial PZT thin films.[28] As shown in Figure 5a-c, the $P_{up}$ and $P_{down}$ domains in bare PZT remain stable over the entire thermal cycle. This helps us establish the baseline for assessing the thermal evolution of domain structures in CIPS, which excludes any influence from the change of underlying PZT.

Figure 5a shows the *in situ* PFM phase images at selected temperatures taken on the 13 nm region of CIPS, which belongs to Region III. From 20 ºC to 175 ºC, the domain structure in CIPS fully conforms to that of the underneath PZT and does not show appreciable change, even though this temperature well exceeds the $T_C$ of bulk CIPS (~42 ºC).[11] Small bubble domains start to emerge at 200 ºC (Figure 5b), clearly illustrating that the sample is approaching $T_C$. At 225 ºC, the CIPS-covered region exhibits noisy PFM responses, the domain structure becomes blurred, and the phase contrast between regions on top of the $P_{up}$ and $P_{down}$ domains of PZT is uniformly suppressed. Similarly, the *in situ* PFM amplitude images (Figure S16a) show that the DWs in the CIPS region are clearly recognizable below 200 ºC and disappear at 225 ºC. We thus conclude that the $T_C$ for the 13 nm CIPS is between 200 ºC and 225 ºC ($T_C \approx 485 \pm 12$ K), which is about 54% higher than the bulk value. After the sample is cooled back down to 20 ºC, the PZT domain structures re-emerge in the CIPS region, indicating that the polar alignment between CIPS and PZT is recovered after the thermal cycle. For comparison, we also perform *in situ* PFM on 14 nm CIPS on Si and Au base layers (Figure S12-15). The results show that the $T_C$ for the CIPS on Si and Au are between 50 °C and 60 °C, which is consistent with previous reports for CIPS flakes and close to the bulk value.[29] It is important to note that the enhanced $T_C$ in thin CIPS exceeds the values for ferroelectric polymer P(VDF-TrFE)[24] and oxide BaTiO$_3$,[25] and approaches that of SrBi$_2$Ta$_2$O$_9$,[26] which makes CIPS highly competitive for developing wearable nanoelectronic and optoelectronic applications.

Figure 5c shows *in situ* PFM phase images at selected temperatures taken on the 55 nm region of CIPS on PZT, which belongs to Region I. At 20 °C, CIPS exhibits randomly distributed domain structures. When heated to 100 ºC, these domains disappear and the overall PFM phase signals become uniformly blurred and noisy, similar to those for the 13 nm region at 225 ºC. The PFM amplitude signal of DWs is clearly recognizable at 20 ºC and becomes fuzzy when heated to 100 ºC (Figure S16b). These results suggest that the 55 nm CIPS flake is already heated above $T_C$. After cooling down to 20 ºC, new random domains emerge, which are different from those before heating, further confirming that the sample has been cycled through $T_C$.[32] This result demonstrates that the $T_C$ enhancement is confined to thin CIPS flakes.

To understand the effect of lattice distortion on $T_C$, we employ the Nudged Elastic Band (NEB) method to calculate the ferroelectric energy barriers for CIPS (see Methods and Supporting Information Section 7). The results reveal an increased energy barrier from 214 meV to 374 meV after setting CIPS on PTO. Fitting the double well energy using the Landau theory shows that the coefficients for the Landau energy expansion correspond to a $T_C$ enhancement of 42% [33, 34] (Supporting Information Section 7). The higher order terms induced by anharmonic couplings[35] may be considered in the future to give a more accurate description of the energy profile, but these terms will not significantly affect the variation of energy barrier induced by PTO. We further perform Monte Carlo simulations of the mean position of Cu in CIPS as a function of temperature (Figure 5d) and set $T_C$ as the temperature at which the mean position of Cu approaches zero. We extract $T_C$ of 400 K for standalone CIPS and 650 K for CIPS on PTO, corresponding to a $T_C$ enhancement of 63%. Considering the slight difference in crystal structures for PTO and PZT, we find the DFT modeling and Monte Carlo simulation results are in reasonable agreement with the experimental result.

**CONCLUSIONS**

In summary, we report controlled domain formation, enhanced ferroelectricity, and tunable piezoelectricity in thin CIPS interfaced with PZT. CIPS polarization is antialigned to that of PZT as the flake thickness is reduced below 25 nm. Thin CIPS flakes on PZT exhibit positive $d_{33}^{\text{CIPS}}$, which increases with decreasing $t_{\text{CIPS}}$, and an enhanced $T_C > 200$ ºC. DFT modeling and Monte Carlo simulations reveal the critical role of interface-mediated lattice coupling in modifying the ferroelectric free energy. Our study provides an effective material strategy to precisely control domain formation and engineer ferroelectricity and piezoelectricity in CIPS, facilitating its implementation in flexible nanoelectronic, optoelectronic, and mechanical applications.

**METHODS**

**Sample preparation**

We deposit epitaxial (001) 50 nm PZT/10 nm LSMO heterostructures on STO substrates using off-axis radio frequency magnetron sputtering. The growth conditions can be found in Ref. [36]. X-ray diffraction *θ-2θ* scans reveal that all PZT films are (001)-oriented with *c*-axis lattice constants of 4.17-4.18 Å (Figure S1), which is consistent with highly strained PZT films.[27] To prepare the Au base layer, we deposit 10 nm Au/2 nm Ti on a $SiO_2$ wafer (Novawafers[@]) using electron beam evaporation. We work with doped Si substrates (Novawafers[@]) with a resistivity of 1-5 mΩ cm.

We mechanically exfoliate CIPS flakes and transfer selected pieces on PZT, Si, and Au base layers using the dry transfer approach at ambient conditions. CIPS flakes are transferred on the pre-patterned domains in PZT after 24 hours of domain writing to ensure that the domain structures are stable and the surface electrostatic conditions are consistent before transfer (Supporting Information Section 5). All sample transfer procedures are conducted at room temperature.

**Conductive AFM and PFM measurements**

The conductive AFM and PFM measurements are carried out using a Bruker Multimode 8 AFM with conductive Pt/Ir-coated probes (Bruker SCM-PIT-V2). For domain writing, we apply a DC bias of ±5 V to the AFM tip and ground the LSMO layer. The selected regions are first poled into the $P_{up}$ state as a uniform background, within which the $P_{down}$ domains are written.

The PFM imaging is conducted with an AC bias voltage close to one of the cantilever's resonant frequencies (310±30 kHz). The driving bias amplitude for all PFM switching and imaging is 300 mV, lower than the coercive voltage of CIPS and PZT. The *in situ* PFM imaging is performed using Bruker's TAC Thermal Application Controller system. The samples are heated on the AFM sample holder and maintained at the target temperature for 30 minutes to achieve thermal equilibrium before imaging. For the PFM phase images shown in main text, the corresponding PFM amplitude images can be found in the Supporting Information.

To measure $d_{33}$, we perform off-resonance PFM ramping measurements. The applied bias is ramped at the sweeping rate of 0.1 V/s. We collect the PFM amplitude at a frequency of 40 kHz, which is well below the cantilever's free-space resonance frequency (~75 kHz). The PFM response is robust and frequency-independent at the frequency range of 20-60 kHz (Supporting Information Figure S7). The applied bias across CIPS and PZT layers is smaller than their coercive voltages in all measurements to avoid polarization switching. The magnitude of $d_{33}$ of a sample is extracted from the slope of PFM amplitude *vs* $V_{bias}$ by the relation: $d_{33} = (S/I) \cdot (\partial A/\partial V_{bias})$. Here $A$ is the output PFM amplitude, $S$ = 54 nm/V is the tip deflection sensitivity, and $I$ = 16 is the system vertical gain. As a control experiment, we characterize the piezoresponse of bare PZT and obtain an average value of $d_{33}^{PZT} = 39 \pm 2$ pm/V, which is comparable with previously reported values for epitaxial PZT thin films.[37] We also perform PFM ramping measurements with different external DC bias superimposed on the AC bias[38] and obtain highly consistent results (Supporting Information Figure S8), confirming that the extracted $d_{33}$ values are not affected by the surface electrostatic charges. More details of the PFM ramping measurement can be found in Supporting Information Section 3.

Each data point of $d_{33}$ shown in Figure 4 is averaged over at least six locations in different regions/flakes with the same thickness. For each location, the measurements have been repeated at least twice to ensure the result is reproducible. For the temperature-dependent measurements (Figure 5), we have performed the *in-situ* PFM imaging twice following the same thermal cycling conditions and obtained consistent results.

**Finite element analysis**

We perform finite element analysis to model the bias voltage distribution across the CIPS/PZT stack. For our sample geometry, we assume a global ground provided by the LSMO layer and a point contact between the conductive AFM tip and the sample surface. Our analysis shows that the fraction of bias voltage across the CIPS layer ($v_{CIPS}$) varies from 0.62 to 0.97 as $t_{CIPS}$ increases from 6 to 248 nm (Supporting Information Section 4).

**DFT calculation**

To obtain the ground state properties of CIPS/PTO interface, we construct the interface supercell based on monolayer CIPS and four layers of PTO. A 2 × 3 supercell of PTO and a rectangular cell of CIPS are adopted for lattice match. PTO is compressed by 21% strain along *a* axis and 5% strain along *b* axis to compensate for lattice mismatch. The core electrons of atoms are eliminated by projector augmented-wave (PAW) potentials[39] as implemented in the Vienna *ab initio* simulation package (VASP).[40] The Perdew-Burke-Ernzerhof (PBE) function[41] is used in the calculation and the spin-orbital coupling is not included. The energy cutoff is 400 eV for structure relaxation and solving the Kohn-Sham equation. We consider a Gamma point sampling in the reciprocal space of the supercell to get ground state energy. A vacuum distance of 15 Å between adjacent layers is used along the periodic direction to avoid spurious interactions. Despite the complexity of ferroelectric perovskite surface due to the interplay of lattice reconstruction,[42] defect states,[43] and charged adsorbates,[44] as well as the anharmonic coupling between the Cu-In polar displacements and the structural distortions in CIPS,[35] our model focusing on the direct lattice coupling between CIPS and PbTiO₃ has successfully captured the enhanced $T_C$ and polarization asymmetry in CIPS interfaced with PZT. Future theoretical efforts involving large-scale *ab initio* or force-field calculations[43] are required to fully account for the complex structural, chemical, and electrostatic interface conditions for this intriguing material system.

**Monte Carlo simulation**

In the Monte Carlo simulation,[45] the length of steps is set to 0.2 Å and the number of steps is 20000. We randomly pick up the direction (positive/negative along *c* direction) of each step with even probability. The acceptance is determined using the Metropolis Hastings algorithm.[45] This process is repeated about 50 times with a fixed starting point, and the average of final positions is used as the result. The entire process is repeated about 100 times to obtain the converged mean values.

**ACKNOWLEDGMENTS**

**Funding:** This work is primarily supported by the National Science Foundation (NSF) through Grant Number DMR-2118828 and EPSCoR RII Track-1: Emergent Quantum Materials and Technologies (EQUATE) Award No. OIA-2044049, and the Nebraska Center for Energy Sciences Research (NCESR). D. Li is supported by the NSF through Grant Number DMR-2118779. L. Yang is supported by the NSF through Grant Number DMR-2124934. The research is performed in part in the Nebraska Nanoscale Facility: National Nanotechnology Coordinated Infrastructure and the Nebraska Center for Materials and Nanoscience, which are supported by the National Science Foundation under Award No. ECCS: 2025298, and the Nebraska Research Initiative. The computational resources are provided by the Extreme Science and Engineering Discovery Environment (XSEDE), which is supported by NSF through Grant Number ACI-1548562. D. Li and L. Yang acknowledge the Texas Advanced Computing Center (TACC) at The University of Texas at Austin for providing HPC resources.

**Author contributions:** X. Hong and K. Wang conceive the project and design the experiments. X. Hong supervises the project. Y. Hao prepares the PZT thin films. K. Wang and Y. Hao carry out structural characterizations of PZT. K. Wang, J. Wang, and H. Anderson prepare the CIPS samples. K. Wang and J. Wang contribute to the conductive AFM and PFM studies. D. Li and L. Yang perform DFT calculations and Monte Carlo modeling. K. Wang and X. Hong write the manuscript. All authors contribute to the results discussion and manuscript preparation.

**Notes:** The authors declare no competing interests.

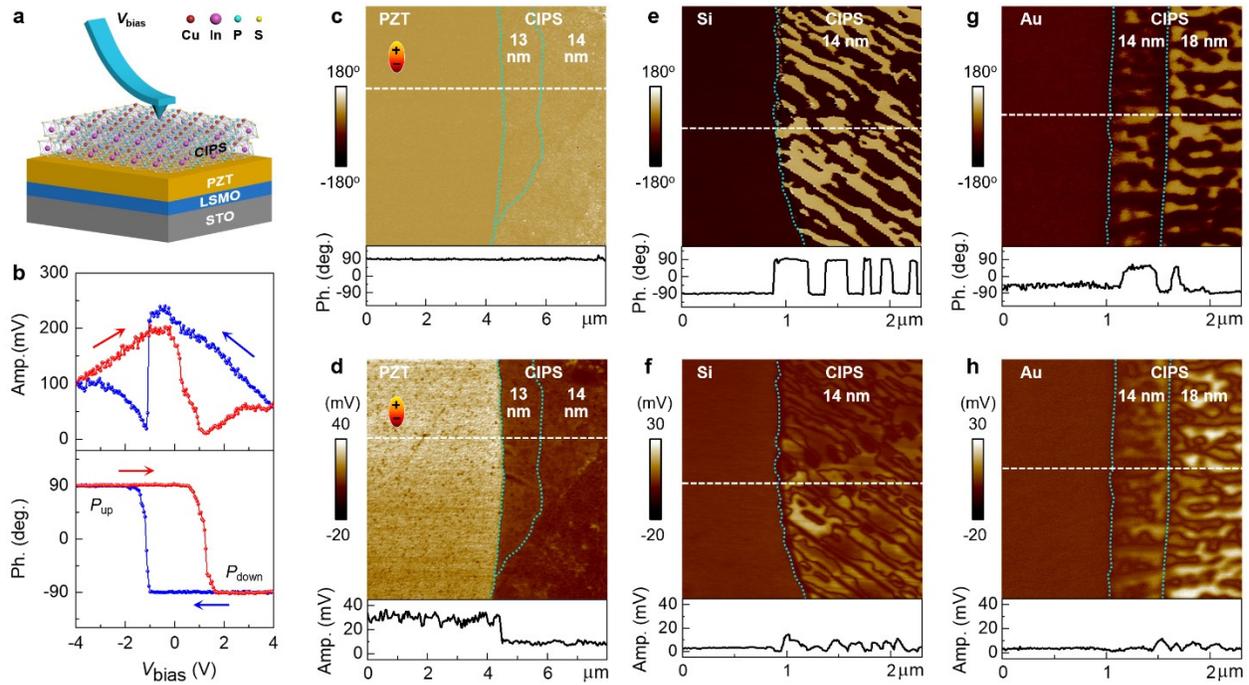

Figure 1. Effect of base layer on domain formation in 14 nm CIPS. (a) Schematic experimental setup. (b) PFM amplitude (top) and phase (bottom) switching hysteresis taken on a 50 nm PZT. (c)-(d) PFM phase (c) and amplitude (d) images taken on 13-14 nm CIPS on $P_{up}$ domain of PZT. (e)-(f) PFM phase (e) and amplitude (f) images taken on 14 nm CIPS on Si. (g)-(h) PFM phase (g) and amplitude (h) images taken on 14-18 nm CIPS on Au. The lower panels in (c)-(h) show the cross-sectional signal profiles along the dashed lines. The dotted lines illustrate the edges of CIPS.

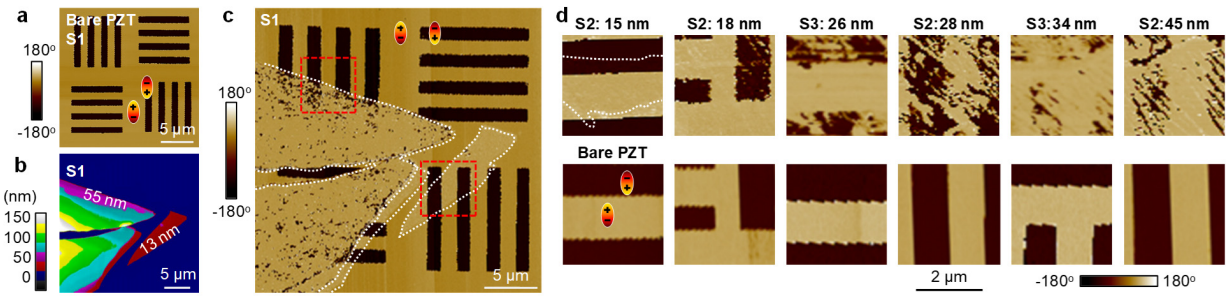

Figure 2. Effect of $t_{CIPS}$ on domain formation in CIPS on PZT. (a) PFM phase image of pre-patterned domains in PZT/LSMO (S1). (b)-(c) AFM topography (b) and PFM phase (c) images of sample S1 with CIPS transferred on top. The CIPS flake thickness is indicated in (b). The dashed boxes outline the regions discussed in Figure 5a,c. (d) PFM phase images taken on samples S2 and S3. Upper panels: CIPS with different thicknesses on pre-patterned domains of PZT. Lower panels: The corresponding domain structures on bare PZT before CIPS transfer. The dotted lines in (c) and (d) outline the CIPS flakes. For bare PZT, the bright (dark) regions correspond to the $P_{up}$ ($P_{down}$) states.

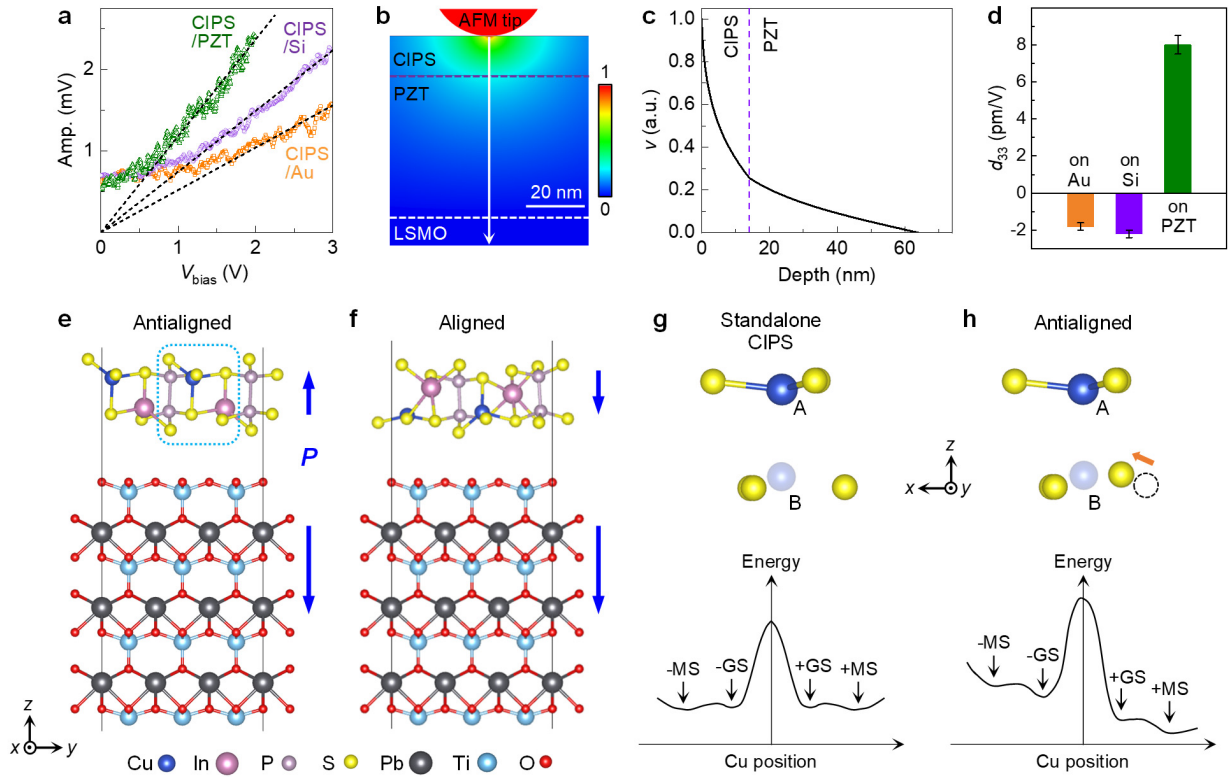

Figure 3. Analysis of $d_{33}$ and polar alignment of CIPS. (a) PFM amplitude *vs* $V_{bias}$ taken on the CIPS samples shown in Figure 1c-h, with linear fits (dashed lines). (b) Simulated distribution of normalized $V_{bias}$ ($v$) across a 14 nm CIPS/50 nm PZT/10 nm LSMO stack, and (c) $v$ along the vertical arrow in (b). The dashed lines indicate the interfaces. (d) Averaged $d_{33}$ for 14 nm CIPS flakes on three types of base layers. (e)-(f) DFT modeling of the crystal structures of relaxed CIPS on PTO frozen in the $P_{down}$ state, with CIPS in the $P_{up}$ (e) and $P_{down}$ (f) states. The arrows indicate the polarization ($P$) directions. (g)-(h) Schematics of sulfur and copper ion arrangements (upper panel) and ferroelectric quadruple-well energy profiles (lower panel) for standalone CIPS (g) and CIPS on $P_{down}$ state of PTO (h). The crystal structure in (h) illustrates the boxed area in (e). The arrow in (h) indicates the shift of sulfur ion relative to its position in (g) (dashed circle). A and B correspond to the Copper ion positions in the $P_{up}$ and $P_{down}$ states, respectively.

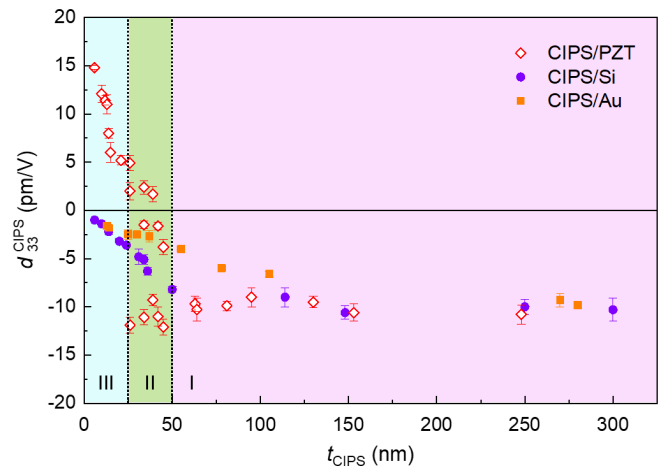

Figure 4. CIPS thickness dependence of $d_{33}^{CIPS}$. $d_{33}^{CIPS}$ *vs* $t_{CIPS}$ for CIPS on different base layers.

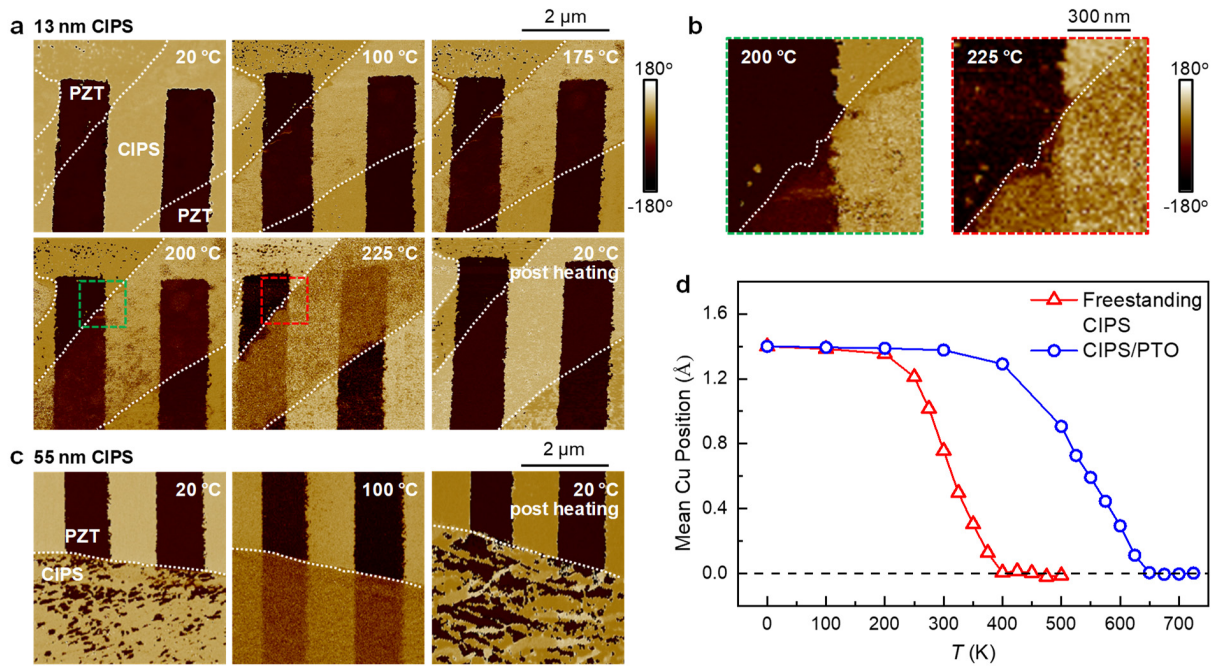

Figure 5. Enhanced $T_C$ for CIPS on PZT. (a)-(c) *In situ* PFM phase images at selected temperatures taken on sample S1 for the 13 nm (a)-(b) and 55 nm (c) CIPS regions outlined in Figure 2(c). The images in (b) are the expanded views of the boxed areas in (a). The dotted lines illustrate the CIPS/PZT boundaries. All images have the same color scale bars. (d) Monte Carlo simulations of mean Cu position at different temperatures.

# Interface-tuning of ferroelectricity and quadruple-well state in CuInP$_2$S$_6$ via ferroelectric oxide (Supporting Information)

Kun Wang[1], Du Li[2], Jia Wang[1], Yifei Hao[1], Hailey Anderson[1], Li Yang[2], and Xia Hong[*,1]

[1] *Department of Physics and Astronomy & Nebraska Center for Materials and Nanoscience, University of Nebraska-Lincoln, Lincoln, Nebraska 68588-0299, USA*

[2] *Department of Physics, Washington University in St. Louis, St. Louis, Missouri 63130-4899, USA*

[*] Email: xia.hong@unl.edu

**Section 1: Characterization of Base Layers**

Figure S1a shows the x-ray diffraction (XRD) *θ-2θ* scan taken on a 50 nm PbZr$_{0.2}$Ti$_{0.8}$O$_3$ (PZT)/10 nm La$_{0.67}$Sr$_{0.33}$MnO$_3$ (LSMO) heterostructure deposited on (001) SrTiO$_3$ substrate, which reveals (001) growth with no impurity phases. The Laue oscillation around the Bragg peak confirms the high crystallinity of the sample. Atomic force microscopy (AFM) measurements of the PZT (Figure S1b), doped Si (Figure S1c), and Au (Figure S1d) base layers show smooth surfaces with the root-mean-square (RMS) roughness ≤ 5 Å.

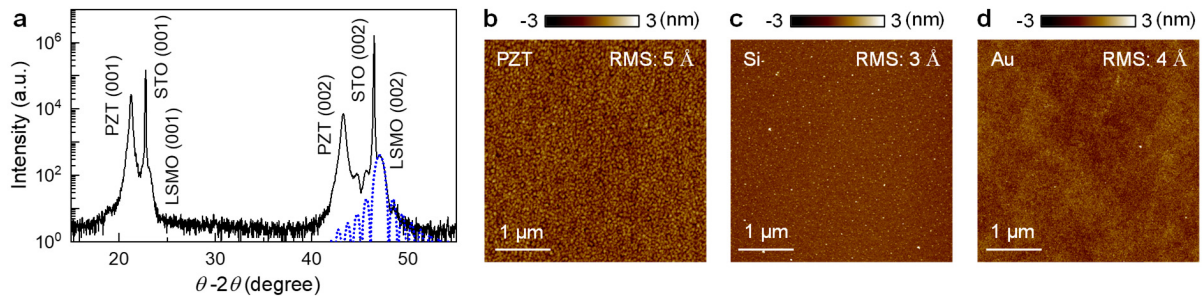

**Figure S1.** Structure and surface characterizations of PZT, Si, and Au base layers. (a) XRD *θ-2θ* scan of a 50 nm PZT/10 nm LSMO deposited on SrTiO$_3$. The blue dotted line is a fit to the Laue oscillation

around the (002) Bragg peak of LSMO. (b)-(d) AFM topography images taken on PZT (b), doped Si (c), and 10 nm Au/2 nm Ti deposited on Si (d).

We write square polarization up ($P_{down}$) and down ($P_{up}$) domains on a bare PZT film using the conductive probe AFM. As shown in the PFM phase (Figure S2a) and amplitude (Figure S2b) images, the as grown film is uniformly polarized in the $P_{up}$ state.

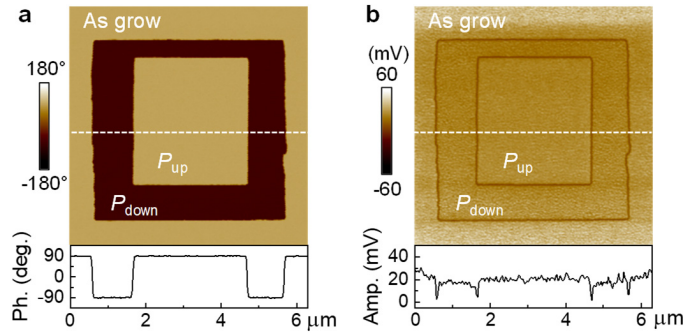

**Figure S2.** PFM characterizations of PZT/LSMO. (a) PFM phase and (d) amplitude images of concentric square domains. The $P_{down}$ ($P_{up}$) dashed box in (a) indicates the domain writing area with $V_{bias}$ = 4 V ($V_{bias}$ = -4 V). The lower panels show the signal profiles along the dashed lines.

Figure S3 shows the AFM topography images of the 14 nm CuInP$_2$S$_6$ (CIPS) flakes on different base layers. The piezoresponse force microscopy (PFM) images of these flakes are shown in the main text Figure 1c-h.

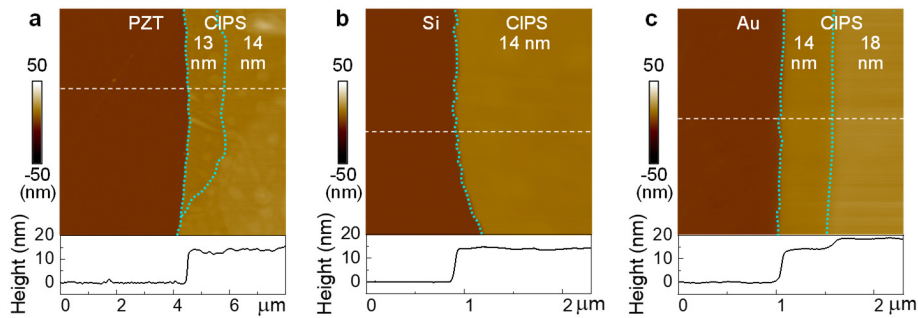

**Figure S3.** AFM measurements of 14 nm CIPS on different base layers. (a)-(c), AFM topography images of CIPS flakes on PZT $P_{up}$ domain (a), Si (b), and Au (c). The lower panels show the cross-sectional signal profiles along the dashed lines. The dotted lines illustrate the edges of CIPS.

**Section 2: CIPS on Prepatterned Domain Structures in PZT**

Using conductive AFM, we create on PZT a series of rectangular $P_{down}$ domains in a uniformly polarized $P_{up}$ background and transfer CIPS flakes on top. Figure S4 shows the PFM amplitude images taken on sample S1 before and after the CIPS transfer.

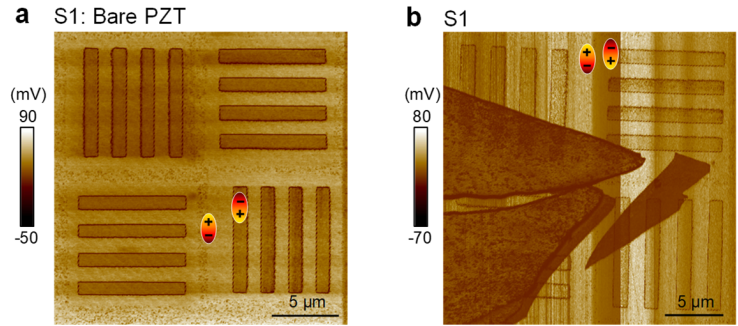

**Figure S4.** Characterization of sample S1. (a) PFM amplitude image of prepatterned domains in bare PZT/LSMO (S1). (b) PFM amplitude image of the same area with CIPS transferred on top. The corresponding AFM topography and PFM phase images are shown in the main text Figure 2a-c.

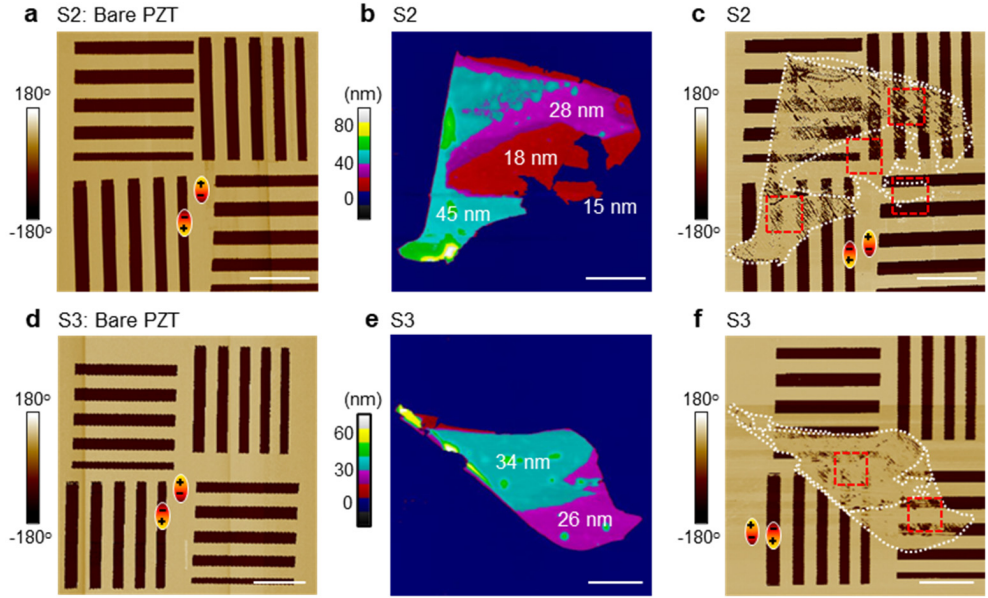

**Figure S5.** Characterization of samples S1 and S2. (d)-(f) AFM and PFM images for CIPS/PZT sample S3. (a) PFM phase image of prepatterned domains in bare PZT/LSMO. (b) AFM and (c) PFM phase images of the same area with CIPS transferred on top. (d)-(f) AFM and PFM images for CIPS/PZT sample S3. (d) PFM phase image of prepatterned domains in bare PZT/LSMO. (e) AFM and (f) PFM

phase images of the same area with CIPS transferred on top. The red dashed boxes in (c) and (f) outline the areas shown in the main text Figure 2d. The scale bars represent 5 µm in all images.

Figure S5a shows the PFM phase image of the prepatterned domain structure taken on bare PZT for sample S2. The AFM and PFM phase images of the same area with a CIPS flake transferred on top are shown in Figure S5b-c, respectively. Figure S5d-f shows the AFM and PFM characterizations of sample S3. Figure S6 shows the PFM amplitude images for the boxed areas in Figure S5c,f. The corresponding phase images are shown in the main text Figure 2d.

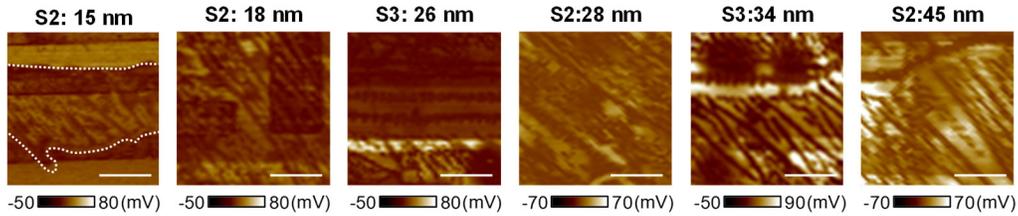

**Figure S6.** PFM amplitude images of the boxed areas in Figure S5c,f. The dotted lines outline the CIPS/PZT boundary. The scale bars represent 2 µm in all images.

**Section 3: Analysis of Piezoelectric Coefficient**

We characterize the effective piezoelectric coefficient $d_{33}$ by performing off-resonance PFM ramping using the Bruker SCM-PIT-V2 probe (spring constant: 3 N/m; tip radius: ~25 nm; free-space resonant frequency: ~75 kHz). The tip deflection sensitivity is calibrated by performing the force ramping before each measurement. The tip deflection sensitivity $S$ = 54 nm/V is extracted from the force curve (deflection *vs* tip height) with $S$ = 1/slope (Figure S7a). We ground the sample bottom electrode, apply an AC bias ($V_{bias}$) at the sweeping rate of 0.1 V/s to the tip, and measure the corresponding piezoresponse. Figure S7b shows the PFM amplitude *vs* $V_{bias}$ taken at frequencies of 10-60 kHz for $V_{bias}$, which are well below the cantilever's free-space resonant frequency. There is less than 10% variation in the slope at the frequency range of 20-60 kHz, showing that the result is robust and frequency-independent. The signals taken at 10 kHz are very noisy. We use 40 kHz for all $d_{33}$ measurements in the main text for the optimized signal-to-noise ratio. From the PFM ramping curves (Figure S7c), we extract $d_{33}$ of 40 ± 0.1 pm/V for the $P_{up}$ state and 37 ± 0.1 pm/V for the $P_{down}$ state of PZT. We use the average value of $d_{33}^{PZT} = 39 ± 2$ pm/V for PZT in analyzing the piezoresponse of CIPS.

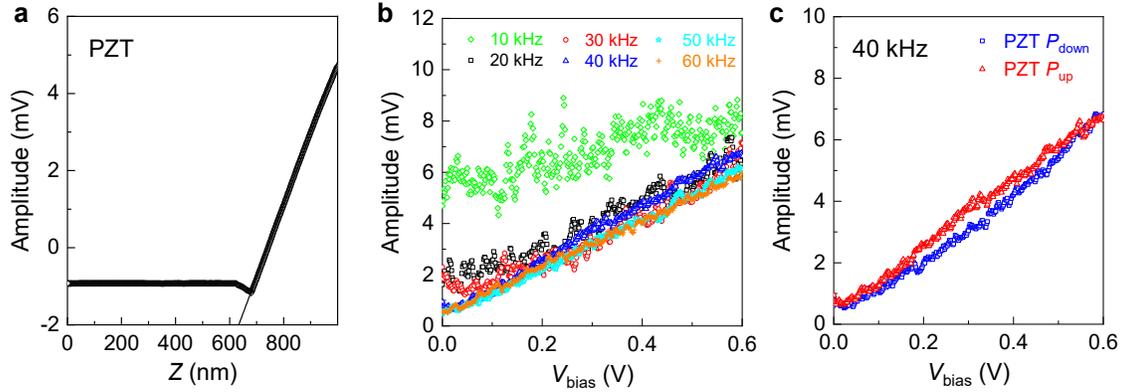

**Figure S7.** Off-resonance PFM ramping measurements of PZT. (a) Deflection amplitude *vs* tip height with a linear fit (solid line). (b) PFM amplitude *vs* $V_{bias}$ at 10-60 kHz taken on the $P_{up}$ domain of PZT. (c) PFM amplitude *vs* $V_{bias}$ at 40 kHz taken on $P_{up}$ and $P_{down}$ domains of PZT.

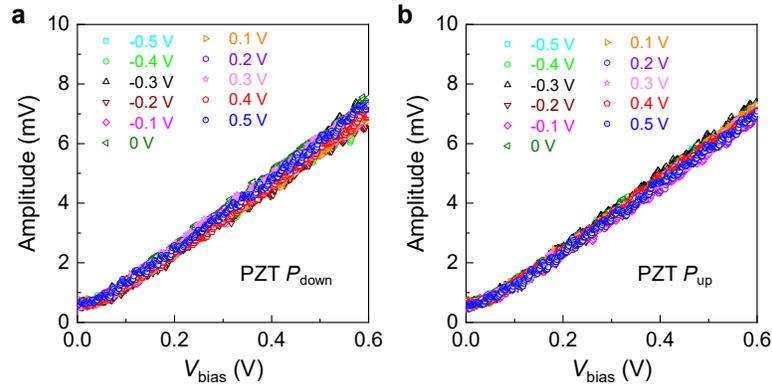

**Figure S8.** Effect of $V_{DC}$ on off-resonance PFM ramping measurements. PFM amplitude *vs* $V_{bias}$ with different $V_{DC}$ from -0.5 V to 0.5 V taken on bare PZT prepatterned into (a) $P_{down}$ and (b) $P_{up}$ domains.

We also examine the effect of the electrostatic force on $d_{33}$ measurements by superimposing a DC bias ($V_{DC}$) when performing PFM ramping measurement.[1] Figure S8 shows the PFM amplitude *vs* $V_{bias}$ taken on bare PZT $P_{down}$ and $P_{up}$ domains with different external $V_{DC}$ from -0.5 to 0.5 V.

The curves are highly consistent and there is less than 15% variation in the slope with different $V_{DC}$. These $V_{DC}$-independent results indicate that our measurements are not affected by the surface electrostatic charges and the obtained piezoresponse of the CIPS/PZT heterostructure is reliable.

To determine the sign of the piezoelectric response of CIPS on Si and Au, we compare their PFM switching hysteresis loops with that of bare PZT. As shown in Figure S9, the PFM phase *vs* $V_{bias}$ loops for CIPS on Si and Au switch in the counterclockwise direction, while that of bare PZT switches in clockwise direction. Since PZT has a positive $d_{33}$, we conclude that $d_{33}$ for CIPS on Si and Au is negative.

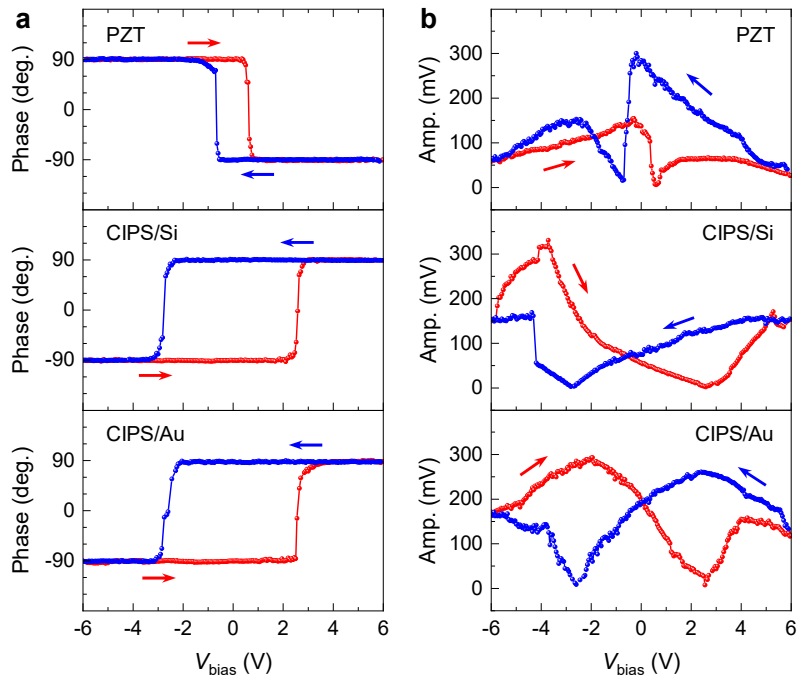

**Figure S9.** PFM switching hysteresis for bare PZT, CIPS on Si, and CIPS on Au. (a)-(b) PFM phase (a) and amplitude (b) switching hysteresis taken on bare PZT (top panels), 42 nm CIPS on Si (middle panels), and 36 nm CIPS on Au (bottom panels). The arrows label the sweeping directions of $V_{bias}$.

## Section 4: Finite Element Analysis of $V_{bias}$ Distribution in CIPS/PZT Stacks

We perform finite element analysis to calculate the bias voltage distribution across the CIPS/PZT stack (main text, Methods). Figure S10 shows the fractional voltage drop across the CIPS layer on 50 nm PZT/10 nm LSMO as a function of CIPS thickness $t_{CIPS}$. The results are used to extract $d_{33}^{CIPS}$ from the net piezoresponse of the CIPS/PZT stack [Equation (1) in the main text]. The finite element analysis results also provide a guide to determine the proper $V_{bias}$ for PFM studies without switching the polarization of the PZT layer.

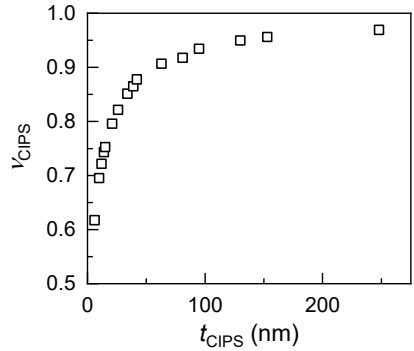

**Figure S10.** Finite element analysis of voltage distribution across CIPS/PZT. The simulated fractional voltage across the CIPS layer ($v_{CIPS}$) *vs* $t_{CIPS}$ for CIPS flakes on 50 nm PZT/10 nm LSMO.

**Section 5: Time-Dependent PFM Measurements of CIPS on PZT**

We perform time-dependent PFM imaging of the domain structure on PZT after CIPS transfer. Figure S11a shows a series of PFM images taken at different time intervals after domain writing at Hour 0. The initial PFM response of the CIPS regions is fuzzy (Hour 1). As the freshly switched domain surface is highly charged and attracts screening charges from the environment, it can cause charge redistribution in CIPS, which affects its PFM response. After 16 hours, the bound charge of PZT is mostly screened,[2] and the domain structure in CIPS fully conforms with that in the underlying PZT. Once settled, the domain structure remains stable for over 11 days, which is the measurement duration.

We also characterize $d_{33}^{CIPS}$ for the 10 nm CIPS region on PZT with time. As shown in Figure S11c, $d_{33}^{CIPS}$ increases from 10.8 pm/V at Hour 1 to 12.1 pm/V at Hour 16 and remains stable afterwards. This value is consistent with $d_{33}^{CIPS} = 12.1 \pm 0.9$ pm/V obtained for the 10 nm CIPS transferred after 24 hours of domain writing (main text, Figure 4). For the PFM studies reported in the main text, we wait for 24 hours after domain writing in PZT at ambient conditions before the CIPS transfer to ensure a stable electrostatic condition of PZT surface.

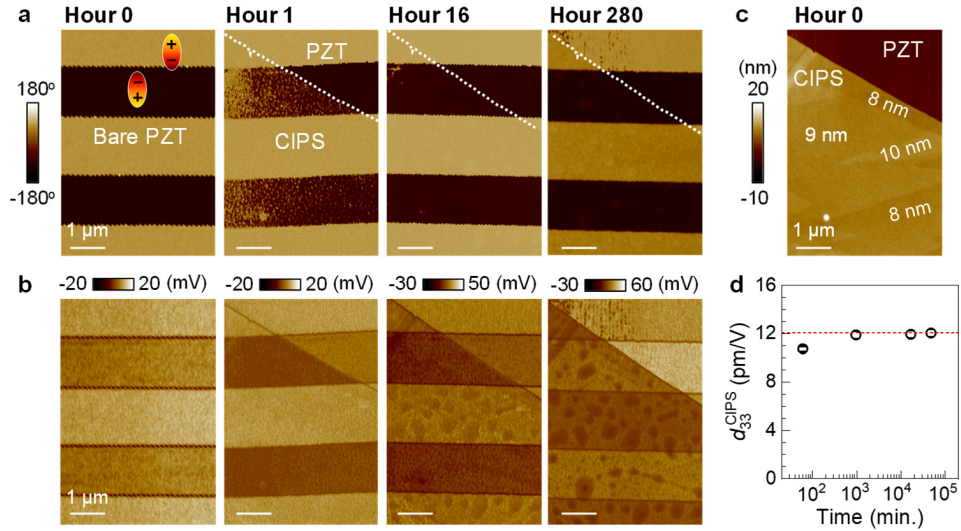

**Figure S11.** Time-dependent PFM imaging of CIPS on PZT. (a)-(b) PFM phase (a) and amplitude (b) images taken on bare PZT right after domain writing (Hour 0) and taken with a CIPS flake transferred on top after 1, 16, and 280 hours of domain writing. The bright (dark) regions in (a) correspond to $P_{up}$ ($P_{down}$) states of PZT. (c) AFM topography image of the sample, with flake thickness indicated. All scale bars represent 1 μm. (d) $d_{33}^{CIPS}$ for the 10 nm CIPS region as a function of time after domain writing. The dashed line marks 12.1 pm/V.

**Section 6: *In Situ* PFM Imaging of CIPS on Si, Au, and PZT**

As a control study, we also examine the Curie temperature ($T_C$) for 14 nm CIPS on Si and Au base layers. We heat the CIPS/Si and CIPS/Au samples at progressive temperatures, wait for 30 minutes after the temperature is stabilized, and then perform *in situ* PFM imaging on the same area. As shown in Figure S12, CIPS on Si exhibits spontaneous stripe domains at room temperature (20 ºC). At 50 ºC, the PFM phase and amplitude signals start to become blurred, suggesting the sample is approaching $T_C$. Above 50 ºC, clear PFM phase contrast is only observed along the step features shown in the sample morphology (indicated by the arrows), while the PFM amplitude is substantially quenched. After cooling back down to 20 °C, new domain structures emerge, further indicating that the CIPS sample has been cycled through $T_C$.

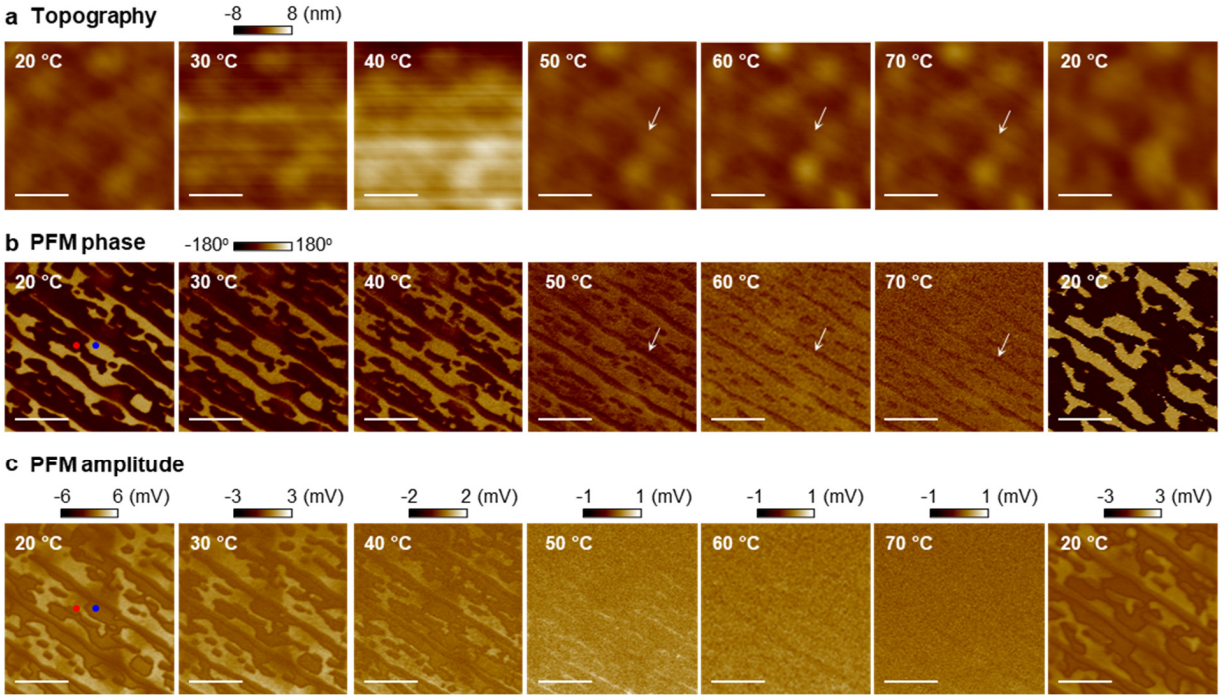

**Figure S12.** *In situ* PFM imaging of CIPS on Si. (a)-(c) *In situ* AFM (a), PFM phase (b), and PFM amplitude (c) images taken on 14 nm CIPS flakes on Si at selected temperatures through thermal cycling. All images in (a) and (b) have the same height and phase signal scales, respectively. The arrows in (a) and (b) indicate the step feature in the topography image and the corresponding PFM phase contrast, respectively. All scale bars represent 500 nm.

Figure S13a-b shows the PFM phase and amplitude signal, respectively, taken on the $P_{down}$ and $P_{up}$ domains of CIPS (labeled positions in Figure S12b-c) at different temperatures. The phase signals for the $P_{up}$ and $P_{down}$ states approach the same value at 60 °C, suggesting the annihilation of domains. The amplitude signal is also quenched at temperatures above 50 °C. We thus conclude that $T_C$ for the 14 nm CIPS on Si is between 50 and 60 °C.

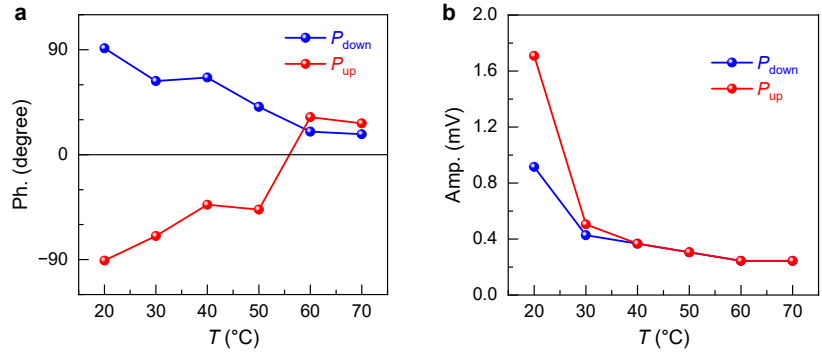

**Figure S13.** Temperature dependence of PFM signal for CIPS on Si. (a)-(b) *In situ* PFM phase (a) and amplitude (b) signal *vs* temperature for the $P_{up}$ and $P_{down}$ states taken at the positions marked by the red and blue dots, respectively, in Figure S12.

Similar temperature evolution of the domain structure has been observed in the 14 nm CIPS on Au. As shown in Figure S14, the domain structures remain stable till 50 °C and become blurred at 60 °C. When heated to 70 °C, the domains completely disappear. After cooling back down to 20 °C, new domain structures emerge, further confirming that the sample has been cycled through $T_C$.

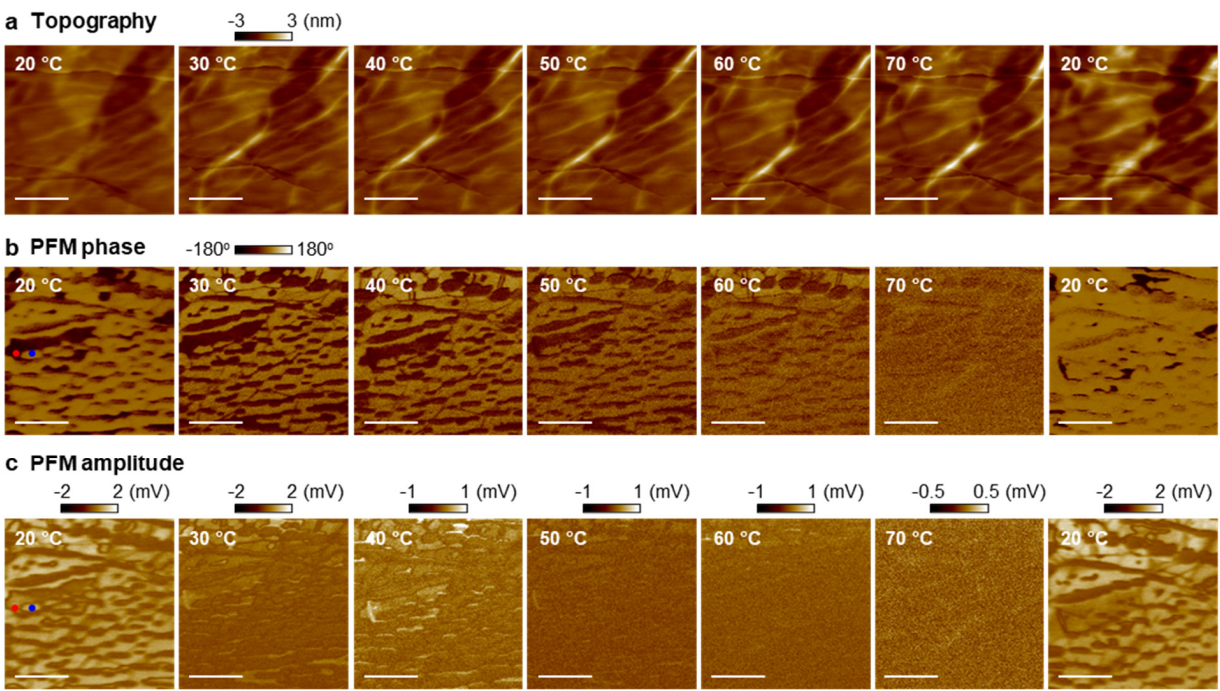

**Figure S14.** *In situ* PFM imaging of CIPS on Au. (a)-(c) *In situ* AFM (a), PFM phase (b), and PFM amplitude (c) images taken on 14 nm CIPS flakes on Au at selective temperatures through the thermal

cycling. All images in (a) and (b) have the same height and phase signal scales, respectively. All scale bars represent 500 nm.

As shown in Figure S15, the PFM phase signal for the $P_{up}$ and $P_{down}$ states for this sample approaches the same value at 60 ºC. We thus conclude that $T_C$ for the 14 nm CIPS on Au is also between 50 and 60 ºC.

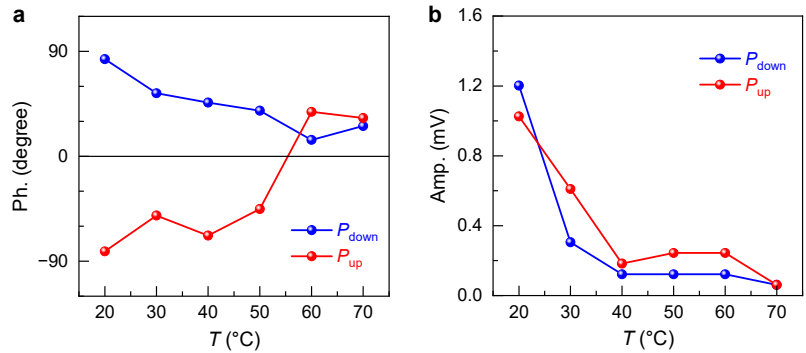

**Figure S15.** Temperature dependence of PFM signal for CIPS on Au. (a)-(b) *In situ* PFM phase (a) and amplitude (b) signal *vs* temperature for the $P_{up}$ and $P_{down}$ states taken at the positions marked by the red and blue dots, respectively, in Figure S14.

Figure S16 shows the *in situ* PFM amplitude images at selected temperatures taken on the 13 nm and 55 nm regions of CIPS in sample S1.

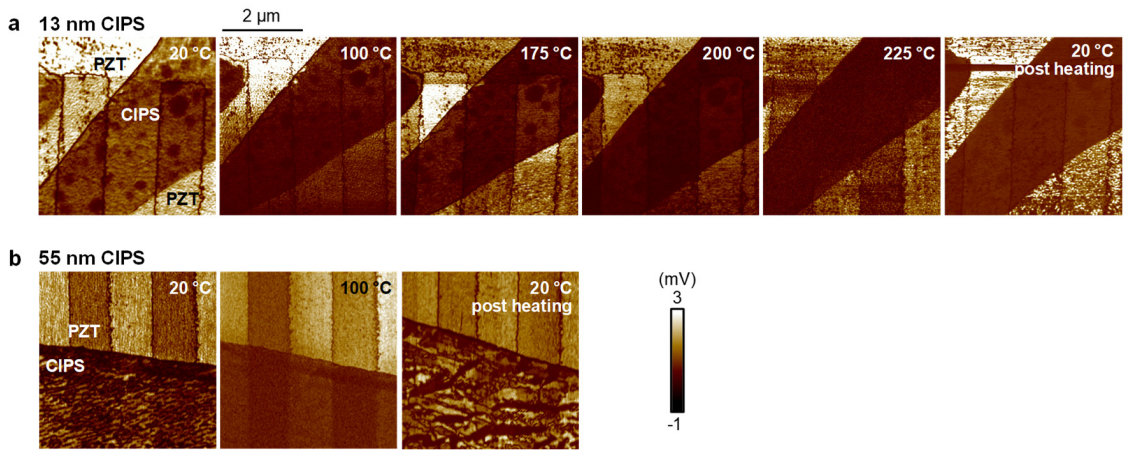

**Figure S16.** *In situ* PFM imaging of CIPS on PZT. (a)-(b) *In situ* PFM amplitude images at selected temperatures taken on sample S1 for the 13 nm (a) and 55 nm (b) CIPS regions. The corresponding

PFM phase images are shown in Figure 5 in the main text. All images have the same length and height scale bars.

**Section 7: Theoretical Modeling**

We calculate the energy barrier for Cu atom displacement using the Nudged Elastic Band (NEB) method. The results are shown in Figure S17. We then fit the free energy double-well within the Landau theory using a 6th-order polynomial: $E = ax^2 + bx^4 + cx^6$, where $x$ is the Cu displacement.[3,4] The fitting parameters are listed in Table S1. The energy barrier per Cu atom is 214 meV for standalone CIPS. In contrast, the energy barrier is enhanced to 374 meV for CIPS on PbTiO$_3$ (PTO), which yields a higher $T_C$ of CIPS. Using Landau theory,[3] we estimate the enhancement of $T_C$ by comparing the ratios of $b/a$ for the two double-well energy fittings. The result shows that $T_C$ for CIPS on PTO is 42% larger than that for standalone CIPS.

We note that the ferroelectric double well for CIPS/PTO is not symmetric. For CIPS on PTO in the $P_{down}$ state, the energy minimum for Cu upward displacement is 25 meV lower than that of downward displacement. It indicates that the $P_{up}$ orientation of CIPS is energetically more favorable for CIPS on the $P_{down}$ state of PTO, or the polarization of CIPS and PTO prefers to be antialigned. Since the 25 meV variation is much smaller than the energy barrier, we can still use the symmetrical polynomial fit in the Monte Carlo simulation.

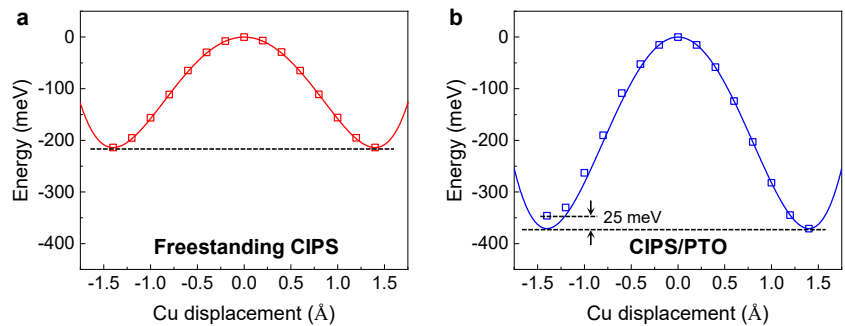

**Figure S17.** DFT modeling of CIPS free energy. (a)-(b) Calculated free energy profile per Cu ion using the NEB method for standalone CIPS (a) and CIPS on PTO in the $P_{down}$ state (b). The solid lines indicate

the polynomial fits to the symmetric energy double well. The positive (negative) Cu displacement corresponds to the upward (downward) direction.

|  | $a$ (meV Å$^{-2}$) | $b$ (meV Å$^{-4}$) | $c$ (meV Å$^{-6}$) | $-b/a$ (Å$^{-2}$) |
|---|---|---|---|---|
| Standalone CIPS | -199.26 | 36.20 | 4.97 | 0.18 |
| CIPS on PTO | -379.70 | 97.71 | 0.29 | 0.26 |

**Table S1.** Fitting parameters for the ferroelectric double-well energy of CIPS.

To understand the microscopic origin of the preference of antialignment between the polarizations of CIPS and PTO, we calculate the crystal structures of CIPS on PTO. As shown in Figure S18, when the polarizations of CIPS and PTO are antialigned, *i.e.*, $P_{up}$ of CIPS on $P_{down}$ of PTO (Figure S18a), all copper ions in the interfacial CIPS layer are aligned in the same horizontal plane. When the polarizations are aligned, *i.e.*, $P_{down}$ of CIPS on $P_{down}$ of PTO (Figure S18b), on the other hand, there is a larger variation in the position of Cu ions along *z*-direction. The corresponding standard deviations of Cu displacement for the $P_{up}$ and $P_{down}$ states of CIPS are 0.01 and 0.15 Å, respectively. The larger variation in Cu ion position causes higher lattice distortion and results in a higher elastic energy cost for the $P_{down}$ state. It is thus preferable for CIPS and PTO to have the polarization antialigned.

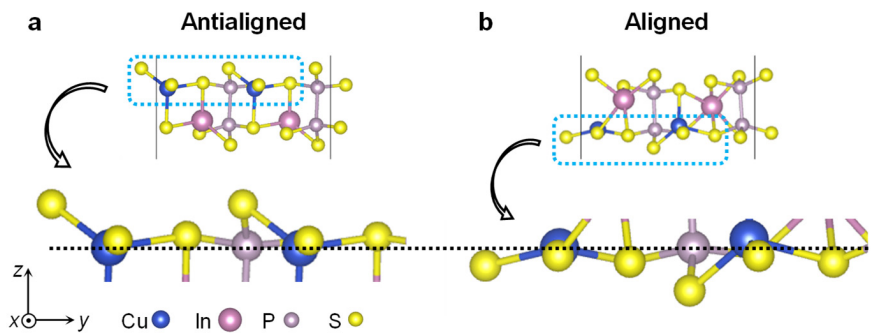

**Figure S18.** DFT modeling of structural distortion of CIPS on PTO. (a)-(b) Calculated crystal structures for CIPS on $P_{down}$ state of PTO with CIPS polarization antialigned (a) and aligned (b) with that of PTO. These are the expanded views of Figure 3e-f in the main text. The bottom PTO layer is not shown for simplicity. Lower panels: Expanded views for the boxed areas in the upper panels. The dotted line serves as a guide to the eye.